\documentclass[aip,sd,amsmath,amssymb,reprint,onecolumn,floatfix]{revtex4-1} 

\usepackage{bm}
\usepackage[mathlines]{lineno}
\usepackage{epstopdf}
 
\usepackage[dvips]{graphicx}
\usepackage{xspace}
\usepackage[usenames,dvipsnames]{xcolor}
\usepackage{array}
\usepackage{ulem}
\usepackage{hyperref}
\usepackage[english]{babel}
\usepackage{lipsum}
\usepackage{booktabs}
\usepackage{colortbl}
\usepackage{xcolor}
\usepackage{xfrac}

\begin{document}

\title{A tunable time-resolved spontaneous Raman spectroscopy setup for probing ultrafast collective excitation and quasiparticle dynamics in quantum materials}
\author{R.~B.~Versteeg}
\email[Corresponding author:]{versteeg@ph2.uni-koeln.de}
\author{J.~Zhu}
\author{P.~Padmanabhan}
\author{C.~Boguschewski}
\author{R.~German} 
\author{M.~Goedecke}
  \affiliation{II. Physikalisches Institut, Universit\"{a}t zu K\"{o}ln, Z\"{u}lpicher Stra{\ss}e 77, D-50937 K\"{o}ln, Germany}
\author{P.~Becker}
\affiliation{Abteilung Kristallographie, Institut f\"{u}r Geologie und Mineralogie, Universit\"{a}t zu K\"{o}ln, Z\"{u}lpicher Stra{\ss}e 49b, D-50674 K\"{o}ln, Germany}  
\author{P.~H.~M.~van Loosdrecht}
\email[Corresponding author:]{pvl@ph2.uni-koeln.de}
  \affiliation{II. Physikalisches Institut, Universit\"{a}t zu K\"{o}ln, Z\"{u}lpicher Stra{\ss}e 77, D-50937 K\"{o}ln, Germany}

\date{\today}

\begin{abstract}
We present a flexible and efficient ultrafast time-resolved spontaneous Raman spectroscopy setup to study collective excitation and quasi-particle dynamics in quantum matter. The setup has a broad energy tuning range extending from the visible to near infrared spectral regions for both the pump excitation and Raman probe pulses. Additionally, the balance between energy and time-resolution can be controlled. A high light collecting efficiency is realized by high numerical aperture collection optics and a high-throughput flexible spectrometer. We demonstrate the functionality of the setup with a study of the zone-center longitudinal optical phonon and hole continuum dynamics in silicon, and discuss the role of the Raman tensor in time-resolved Raman scattering. In addition, we show evidence for unequal phonon softening rates at different high symmetry points in the Brillouin zone of silicon by means of detecting pump-induced changes in the two-phonon overtone spectrum. Demagnetization dynamics in the helimagnet Cu$_2$OSeO$_3$ is studied by observing softening and broadening of a magnon after photo-excitation, underlining the unique power of measuring transient dynamics in the frequency domain, and the feasibility to study phase transitions in quantum matter. 
\end{abstract}

\maketitle

\section{Introduction}
The last decade has seen a surge of experiments where light is exploited as a strong external stimulus to manipulate quantum matter. \cite{basov2017,zhang2014,gandolfi2017} The light-matter interaction can be a fully coherent process as in the case of the Floquet state in photon-dressed materials, \cite{gedik2013} or drive quantum matter into a strongly non-thermodynamic state by disturbance of the balance between electronic, orbital, spin, and lattice degrees of freedom. Seminal cases of the latter phenomenon include the photo-induced insulator-metal transition in VO$_2$, \cite{cavalleri2001} optical control of colossal magnetoresistivity in manganites, \cite{rini2007} and transient signatures of photo-induced superconductivity in a stripe-ordered cuprate. \cite{fausti2011} 

Integral in the description of the emanating dynamics in the optically driven non-equilibrium state is the creation and annihilation of electronic quasiparticles, collective excitations associated with the various degrees of freedom, and their interaction. Their transient annihilation and creation may for instance induce symmetry changes across photo-induced phase transitions, \cite{boschini2018} and dictates nonequilibrium temperatures, \cite{maldonado2017} while their interaction underlies the coupling, and therefore the equilibration between the various degrees of freedom. \cite{turgut2016} A scala of partially complimentary pump-probe techniques have been developed to study quasiparticle and collective excitation dynamics, and their interaction.
For instance, the coupling between low-energy collective excitations, and high-energy electronic excitations may be mapped into the time-domain, and quantified, through all-optical coherent fluctuation spectroscopy. \cite{mansart2013,mann2015} 
A direct view on momentum-dependent electron and hole quasiparticle population dynamics is provided by means of the matured technique of time-resolved angle-resolved photoemission spectroscopy. \cite{sobota2012,boschini2018} The situation for direct probing of collective excitation populations is perhaps more ominous. \cite{abbamonte2013} Rather novel photon-probes of dynamics of collective excitations such as phonons, magnons, and orbital excitations are time-resolved diffuse x-ray scattering, \cite{trigo2013fourier} and time-resolved resonant inelastic x-ray scattering. \cite{dean2016ultrafast} Plasmon dynamics in quantum materials has for instance succesfully been probed with femtosecond electron energy loss spectroscopy. \cite{carbone2009,piazza2014} However, a quantification of occupation numbers and corresponding effective temperatures with these techniques still seems a stretch.

Time-resolved Raman scattering is a reasonably well-established probe for vibrational dynamics in organic materials, \cite{iwata1993,hamaguchi1994,uesugi1997,sahoo2011,kruglik2011} and phonon population dynamics in semiconductor materials \cite{kash1985,oberli1987,zhu2017} and carbon allotropes. \cite{song2008,yan2009,kang2010,yang2017novel} With this technique the transient evolution of the spontaneous Raman spectrum after photo-excitation is studied. \cite{faustibook} The pump excitation mechanism only needs to induce incoherent dynamics. There are thus no special prerequisites to the pulse duration, as is the case with coherent pump-probe techniques. \cite{mansart2013,mann2015} In the time-domain spectra transient changes in excitation energies, line-widths, and scattering intensities of Raman-active excitations can be obtained. This has the advantage over coherent excitation techniques that one can follow the true time evolution of the system through its incoherent response, rather than the time evolution of coherent excitations. Through detailed balance of the anti-Stokes to Stokes scattering intensity ratio the temporal evolution of selected effective mode temperatures can be directly determined,\cite{faustibook,compaan1984} as in contrast to techniques relying on a comparison of the dynamical response with the thermodynamic temperature evolution. This makes time-resolved spontaneous Raman spectroscopy a truly unique technique in the study of ultrafast processes, and for instance allows testing of the validity, and limitations of widely applied multiple effective temperature models. \cite{waldecker2016,maldonado2017} An interesting situation is expected on the shortest time-scales, where truly non-thermal dynamics is present and the fluctuation dissipation theorem may not hold anymore. Phase transitions can be optically induced when a quantum material is driven far out-of-equilibrium. The associated symmetry changes across the phase transition can be deduced from the changes in selection rules of the transient Raman spectrum. \cite{fausti2009} Since the transient excitation dynamics is directly measured in the frequency domain, the renormalization of excitation energies and linewidths across the phase transition (or partial melting of a phase) may be measured with large accuracy. This should be compared and contrasted with time-domain techniques which rely on the detection of a frequency chirp in the coherent excitation dynamics. \cite{misochko2004}

Major factors to the limited use of time-resolved spontaneous Raman spectroscopy in the study of ultrafast dynamics in quantum matter are energy-time resolution limitations, the relevant excitation energy scales, \cite{basov2011} and low inelastic light scattering cross sections. In time-resolved Raman spectroscopy the time resolution and spectral resolving power are Fourier transform related. This asks for a proper choice of design parameters as we recapitulate in this article. The Fourier transform limit can in principle be overcome with the femtosecond stimulated variant of the technique. \cite{kukura2007femtosecond} A recent success with time-resolved femtosecond stimulated Raman spectroscopy (tr-FSRS) is the observation of a modification of the Heisenberg exchange in an antiferromagnet on the femtosecond timescale. \cite{batignani2015probing} In FSRS the stimulated gain and loss intensities both have contributions from anti-Stokes and Stokes processes. The gain and loss ratio therefore cannot be easily interpreted in terms of an occupation number or equivalent temperature. \cite{harbola2013} Time-resolved spontaneous Raman therefore stays an appealing technique to study quasiparticle and collective excitation population dynamics, and optimization of signal detection remains a central issue. Recent advances in high repetition rate amplified laser systems, tunable light sources, and detection techniques can nowadays bring the sensitivity of time-resolved spontaneous Raman spectroscopy close to that encountered in a continuous wave laser based steady-state spontaneous Raman scattering experiment, even without relying on resonances. A resurgence of the technique as a tool to probe quasiparticle population dynamics, photo-induced symmetry changes, and energy and momentum-transfer in quantum materials is thus foreseen.

In this paper we present a flexible and efficient ultrafast time-resolved spontaneous Raman spectroscopy setup to study collective excitation and quasiparticle dynamics in quantum materials. The setup has a broad energy tuning range extending from the visible to near infrared spectral regions for both the pump and probe pulse energies. It thereby allows to selectively excite materials, and to probe under off- and on-resonant Raman conditions. The balance between energy and time-resolution can be controlled, allowing few picosecond time-resolution studies with an energy resolution down to $\Delta\nu$\,$\sim$\,$7$\,cm$^{-1}$ or femtosecond studies with $\Delta\nu$\,$\sim$\,$50$\,cm$^{-1}$ energy resolution. A high light collecting efficiency is realized by high numerical aperture collection optics, and a high-throughput flexible spectrometer. We demonstrate the functionality of the setup with three different case studies. We first focus on the zone-center longitudinal optical phonon (LO-phonon), and hole continuum dynamics in photo-excited silicon, and discuss the role of the Raman tensor by simultaneously tracking the Stokes and anti-Stokes spectra. Changes in the Raman tensor resulting from photo-induced symmetry changes have been discussed previously. \cite{fausti2009} Here we show that changes in the Raman tensor can additionally appear due to resonance effects. This can be used as an alternative probe to track electronic population dynamics. Higher order Raman processes can serve as an all-optical momentum dependent probe. We show evidence for dissimilar electron-phonon scattering rates at different high symmetry points in the Brillouin zone of silicon by detecting pump-induced changes in the two-phonon overtone spectrum. This measurement underscores the unique resolving power of measuring transient changes in excitation energies in the frequency domain. Demagnetization dynamics of helimagnetic order in the chiral insulator Cu$_2$OSeO$_3$ is studied by observing softening and broadening of a magnon excitation after photo-excitation. We thereby illustrate the technique's feasibility to probe photo-induced phenomena in quantum materials, such as melting of phases, and photo-inducing phase transitions. With the high stability and sensitivity of the described setup new avenues in ultrafast dynamical studies of quantum matter are foreseen. These include the tracking of order parameter evolution and detection of symmetry changes across photo-induced phase transitions, the qualitative determination of energy and angular momentum transfer rates in the relaxation of non-equilibrium states, and time- and momentum-resolved scattering.

\section{The time-resolved spontaneous Raman spectroscopy setup}
\subsection{Design considerations}
\begin{figure}[h!]
\center
\includegraphics[width=2.375in]{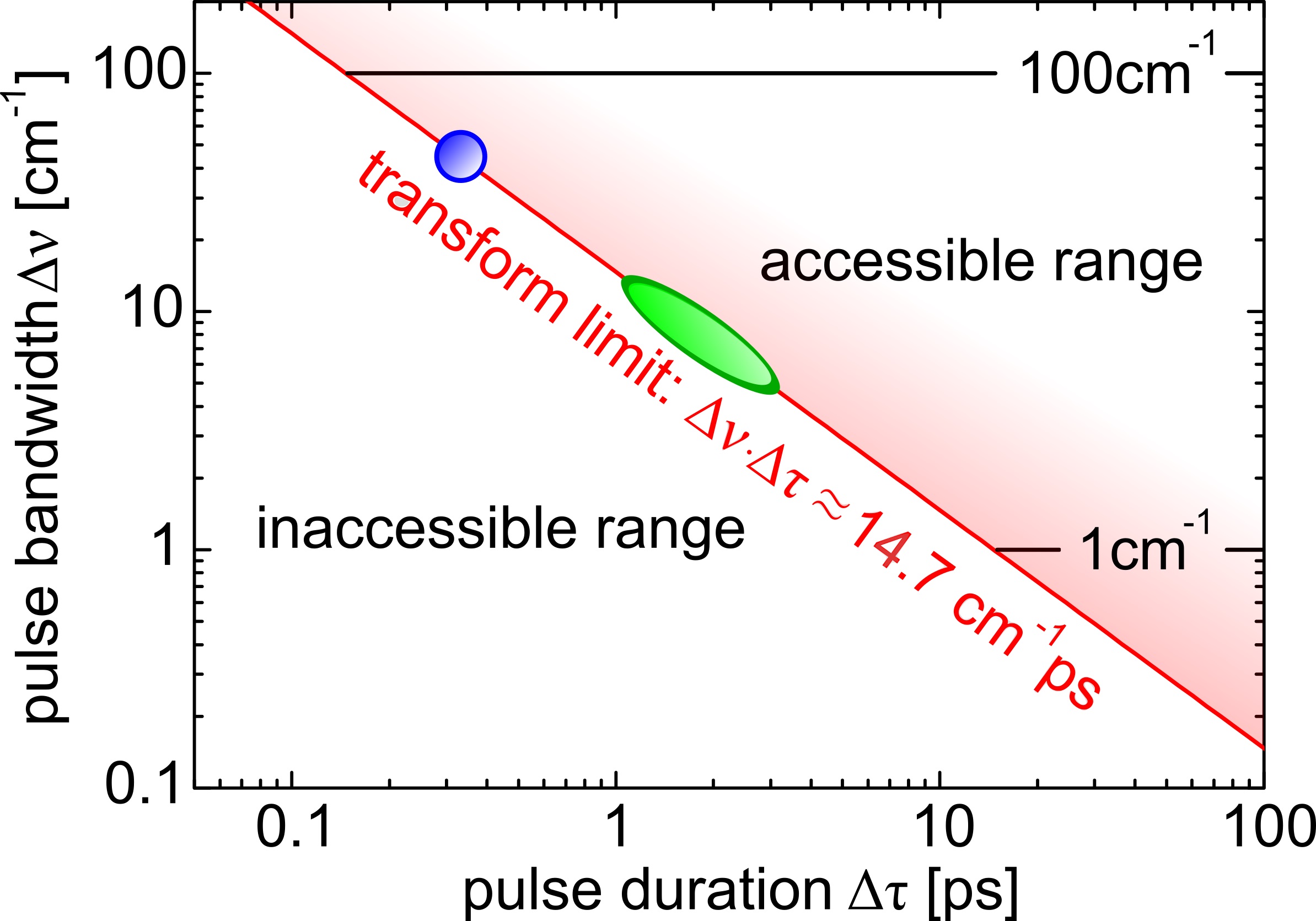}
\caption{The Fourier transform relation $\Delta \nu\Delta \tau$\,$\approx$\,$14.7$\,cm$^{-1}$ps between pulse duration $\Delta \tau$ and pulse bandwidth $\Delta \nu$ puts a limit to the experimentally accessible energy and temporal dynamics range. The energy resolution region of interest for time-resolved spontaneous Raman studies of quantum materials is put between $1$\,cm$^{-1}$ and $100$\,cm$^{-1}$. This leads to a few tens of picoseconds, to a few hundred of femtoseconds temporal resolution. The green ellipse indicates where the system operates in ps-mode. The time, and energy resolution can be controlled by detuning the SHBC, or by a pulse shaper. The blue dot is where the system operates when probe light is generated with the FS-OPA.}
\label{fig:transformlimit}
\end{figure}

Typical design parameters for a time-resolved spontaneous Raman setup include energy resolution, time-resolution, the laser linewidth, and laser repetition rate. \cite{faustibook} Other design factors include tunability of the pump excitation and Raman probe light, and suitable schemes to filter the Raman scattered light from the elastically scattered light. The energy- and time-resolution parameters are of special importance in time-resolved spontaneous Raman spectroscopy. For a transform-limited Gaussian pulse the bandwidth-time relation is $\Delta \nu \Delta \tau$\,$\approx$\,$14.7$\,cm$^{-1}$ps, with $\Delta \nu$ the frequency (energy) bandwidth of a pulse, and $\Delta \tau$ the pulse duration. Here $\Delta \nu$ and $\Delta \tau$ refer to the full width at half maximum (FWHM). A trade-off in time and energy resolution is thus inherent to the technique, and ideally a setup should allow control of energy and temporal resolution. The experimentally accessible region, limited by the Fourier transform relation, is indicated in Fig.\,\ref{fig:transformlimit}. In materials with only a few Raman-active modes, such as IV and III-V semiconductors, \cite{kash1985,kttsen2006} or well-separated high-energy modes in for instance graphite allotrope materials, \cite{song2008,yan2009,kang2010} a pulse bandwidth of $\Delta \nu$\,$\approx$\,$100$\,cm$^{-1}$ (FWHM) suffices to resolve individual modes, and thus sub-ps time-resolution can be realized (see Fig.\,\ref{fig:transformlimit}). This energy resolution however generally isn't sufficient to study dynamics of collective excitations in quantum materials such as magnons \cite{fleuryloudon1968} and phonons, or the more complex types such as electromagnons, \cite{rovillain2012} Cooper-pair breaking, \cite{saichu2009} and charge-density wave modes \cite{sugai2006} since these modes generally are low energy excitations with $\Delta$E$\,<120$\,meV. This energy scale corresponds to Raman shifts of $\sim 1000$\,cm$^{-1}$ and lower, and thus requires a higher spectral resolving power. This holds especially in the lowest energy region ($\Omega$\,$<$\,$100$\,cm$^{-1}$) where for instance soft modes in ferroelectrics are observed.\cite{shigenari2015raman} In this region stray light from spectrally broad laser pulses can hinder the observation of the modes of interest. Another case where high resolving power is necessary are materials of low crystallographic symmetry where more interesting modes of electronic and magnetic origin might overlap with a multitude of phonon modes.

In spontaneous Raman spectroscopy the inelastically scattered light needs to be separated from the Raman excitation light. For higher energy modes such as vibrational excitations in carbon-based materials this can for instance be realized with notch filters. When tunability in probe wavelength is of importance a triple subtractive Raman spectrometer is still favored. In time-resolved spontaneous Raman spectroscopy an additional complication appears since the pump-excitation, and the pump-induced Raman spectrum need to be rejected. Different filtering schemes have been applied for this. \cite{faustibook} In a degenerate pump-probe experiment polarization optics are used to reject the pump-induced elastic, and inelastic scattered light. The downside is that only the parallel Raman polarization geometry can be studied. \cite{fausti2009} In addition, unwanted polarization leakage also ends up in the Raman spectrum, and the lowest energy excitations are difficult to access. In a two-color setup however, the pump-induced elastic (and inelastic) scattering can be conveniently rejected by spectral filtering. \cite{uesugi1997,faustibook}
	
Amplified laser systems allow for tunable two-color experiments, as opposed to MHz-oscillator experiments which are limited to the fundamental and double wavelength for the pump and probe beams. In addition, with an amplifier the problem of average heating is avoided. For low repetition rate systems the average laser power can however bring challenges. The Raman intensity scales with average laser power. For detectable average laser powers with low repetition rate systems the pulse peak intensities can thus easily get too large, which may lead to nonlinear effects, and sample damage. The ideal situation between these laser system limits thus exists in the form of high repetition rate amplifiers. \cite{uesugi1997,faustibook}

\subsection{System overview}
\begin{figure}[h!]
  \includegraphics[width=5.375in]{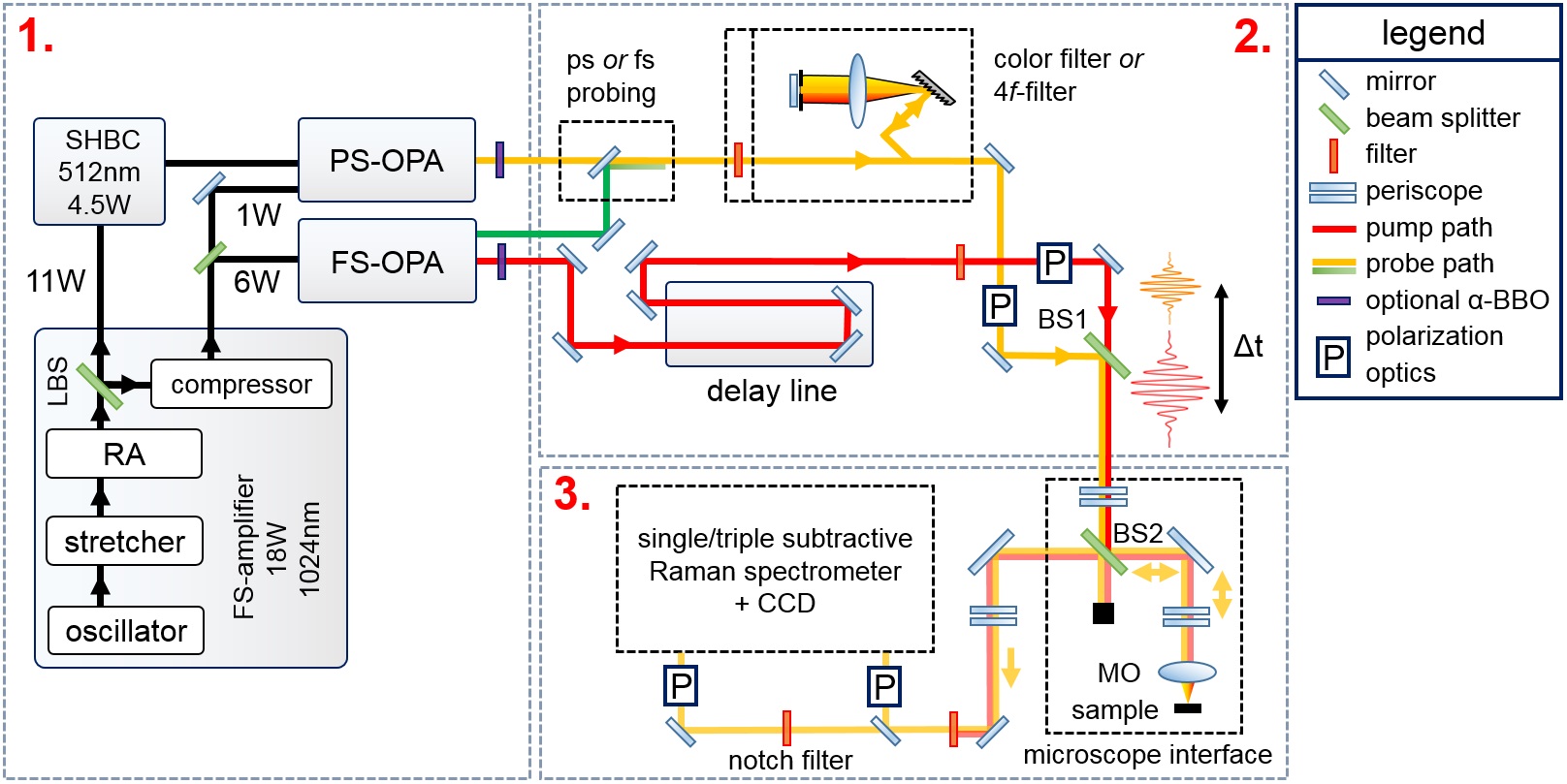}
  \caption{The femtosecond-picosecond time-resolved spontaneous Raman setup. Box 1.) Laser system and optical parametric amplifiers. Box 2.) Table optics and delay line. Box 3.) The confocal Raman microscope interface and detection system. See the text for details.}
\label{fig:setupfig}
\end{figure}

The time-resolved Raman system consist of three main parts: 1.) the amplified laser system and optical parametric amplifiers to generate pulses for selective pumping, and narrow-bandwidth pulses for Raman probing, 2.) the table optics for pulse cleaning, polarization control and the delay line 3.) the confocal Raman microscopy interface, the high efficiency spectrometer and charge-coupled device detector. A layout of the setup is shown in Fig.\,\ref{fig:setupfig}. 

\subsubsection{Amplified laser system \& selective pump excitation and tunable Raman probe}
A chirped pulse amplification based laser system (\textsc{LightConversion Pharos}) was selected for the setup. The \textsc{Pharos} laser system consists of a Kerr-lensing mode locked oscillator module, a regenerative amplifier (RA), and stretcher-compressor units which are all embedded in a compact module. The laser uses diode-pumped Yb:KGW (Ytterbium-doped potassium gadolinium tungstate) as the active medium. The emitted pulses have $\lambda _c$\,=\,$1024$\,nm central wavelength. An internal pulse picker allows to set the repetition rate up to $f_{\rm max}$\,=\,$100$\,kHz. After the regenerative amplification stage the $1024$\,nm stretched pulse beam is split into two beams. One beam is directly emitted as an $11$\,W output of non-compressed highly chirped $150$\,ps long pulses. The other beam is compressed, resulting in a $7$\,W average power beam of $0.3$\,ps duration pulses.  

The uncompressed $11$\,W beam is routed to a second harmonic bandwidth compressor (\textsc{LightConversion SHBC}). \cite{raoult1998} Inside the SHBC, a 1:1 beamsplitter divides the beam. A high-intensity grating gives the split beams an opposite chirp, after which the beams are overlapped on an $\alpha$-BBO crystal. The SHBC converts the $1024$\,nm pulses of full-width at half maximum (FWHM) $\Delta\nu$\,$\approx$\,$50$\,cm$^{-1}$, into $512$\,nm transform limited pulses with $\Delta\nu$\,$\approx$\,$10$\,cm$^{-1}$ FWHM and $\Delta \tau$\,$\approx$\,$1.5.$\,ps temporal width. The SHBC output can directly be used as Raman probe light, or used to pump a three stage white-light seeded picosecond optical parametric amplifier (\textsc{LightConversion PS-OPA}). The white light seed is generated with $1$\,W of $0.3$\,ps $1024$\,nm compressed pulses. The PS-OPA can continuously tune the Raman probe wavelength from $630$\,nm to $950$\,nm. The signal and idler of the PS-OPA output can be externally doubled to tune the probe wavelength in the range $320$\,-\,$600$\,nm. 

A total power of P\,=\,$6$\,W of the $0.3$\,ps $1024$nm compressed beam is routed to a double-pass white-light seeded optical parametric amplifier (\textsc{LightConversion Orpheus FS-OPA}) to generate pump pulses. The signal wavelength is continuously tunable from $620$\,nm to $1000$\,nm, where the idler runs from $1060$\,nm to $2500$\,nm. An external $\alpha$-BBO crystal can be used to extend the pump wavelength range from $320$ to $600$\,nm.

The time-resolved system operates in the green oval region in the ($\Delta \nu$,$\Delta \tau$)-plane of Fig.\,\ref{fig:transformlimit} when probing is realized with the SHBC or PS-OPA output. High time-resolution, with lower energy resolving powers can be realized by only working with the FS-OPA. The FS-OPA's signal output is in this case still used for pumping. The residual output of the OPA's $0.3$\,ps, $\lambda _c$\,=\,$512$\,nm pump beam is used for probing. For the FS-OPA based operation mode the system works in the blue sphere in the ($\Delta \nu$,$\Delta \tau$)-plane.

\subsubsection{Table optics}
Narrow-band laser line filters are placed in the pump beam to remove unwanted spectral components which may otherwise lead to spurious signals in the Raman spectra. The polarization state of the pump beam can be controlled by a Berek compensator. For the probe beam the unwanted spectral components are either removed by laser line filters or a folded grating-based pulse shaper, \cite{kawashima1995femtosecond} where the latter results in the cleanest pulse, as required for probing small Raman shifts ($\Omega$\,$<$\,$100$\,cm$^{-1}$). In addition, the pulse shaper allows to narrow down the spectral pulse bandwidth at least to about $\Delta \nu$\,$\approx 7$\,cm$^{-1}$ FWHM at the expense of time-resolution and average probing power. A Berek compensator is used to control the polarization state of the pump beam.

The temporal delay between the pump and probe pulse is controlled via a mechanical delay line of $30$\,cm length (\textsc{Physikalische Instrumente M-531.EC}) placed in the pump beam path. The step size resolution is $\sim$\,$0.5$\,$\mu$m, corresponding to a time-resolution of approximately $\sim$\,$3$\,fs. The time-resolution is thus effectively determined by the pump-probe pulse cross-correlation. A silver-coated retro-reflector is mounted on the delay line. The pump and probe beams are made collinear before the Raman microscopy interface with beam combiner BS1. 

\subsubsection{Raman interface, spectrometer and CCD-detector}
The collinearly propagating pump and probe beams enter the Raman microscopy interface via a periscope. The collinear pump and probe beams are reflected from beam splitter BS2 (R/T\,=\,0.2/0.8) into a high numerical aperture microscope objective MO (\textsc{Olympus LMPLFLN}-series long working distance objectives are used, where the NA=$0.4$, $20\times$, working distance WD=\,$12$\,mm is the standard choice). The backscattered light is collected by the microscope objective, transmitted through beam splitter BS2, and focused on the spectrometer entrance slit.

The maximum pump pulse energy lies around $\sim$\,$150$\,nJ at the sample position for $650$\,nm pump excitation. The maximum probe pulse energy lies around $\sim$\,$20$\,nJ at the sample position for $512$\,nm probe excitation when the grating-based pulse shaper is used. Since the Raman scattering efficiency scales with incident laser intensity, a large average incident probe power is preferred. For diffraction-limited pump-probe spot sizes this however results in a few challenges. Pulse fluences get too large, and samples damage easily. This can be solved by defocussing the microscope objective. The divergence of the scattered light in this case is corrected with a telescope system placed after BS2. This allows working with spot sizes of about $\sim$ \,$1$\,$\mu$m to $\sim$\,$100$\,$\mu$m diameter. The samples can be placed in a \textsc{Janis ST500} coldfinger cryostat with small working distance (about $2$\,mm).

A successful implementation of time-resolved Raman spectroscopy significantly hinges on the efficiency of the spectrometer. Raman-scattered light is an order $\mathcal{O}$\,$\sim$\,$10^{-8}$ weaker compared to the incident excitation light. This necessitates a high stray light rejection. We use a cascade of three Czerny-Turner \cite{richardsongrating} imaging spectrographs (\textsc{TriVista 555, S\&I GmbH}) which are operated in subtractive mode (high stray light rejection by the first two spectrometers). Different sets of holographic, and ruled gratings allow to maximize the scattered light detection efficiency. Silver-coated optical elements ensure a high throughput. The last stage (spectrometer stage) has a second entrance port. This allows to bypass the subtractive stage of the spectrometer. In this case a notch filter is used to block the elastically scattered light. The use of a notch filter can provide up to a factor $2$ increase in throughput efficiency with respect to the subtractive filtering. A color filter may additionally be used to suppress the scattered pump light.

The used imaging spectrometer \cite{lerner2006} contains toroidal mirrors which corrects for the astigmatism present in a spectrometer containing spherical mirrors. In the toroidal case a smaller image size in the lateral direction is thus produced on the charge-coupled device (CCD) chip. This has the advantage that less CCD-rows need to be binned to integrate the scattered light spectrum. This significantly reduces the appearance of occasional spikes from cosmic events in the Raman spectrum, and reduces the CCD-readout noise with respect to an ordinary spectrometer. The low spherical aberration of the spectrometer makes that one can resort to the use of pseudo-confocal microscopy. \cite{williams1994} In this case the entrance slit rejects out-of-focus light in the horizontal direction, whereas the CCD-binning is used to digitally reject out-of-focus light in the vertical direction from the Raman spectrum. By resorting to pseudo-confocality the use of an opto-mechanical confocal system is avoided. In the latter case a secondary focus point through a mechanical pinhole before the spectrometer is used to obtain confocality. This would however lead to reflection losses for the scattered light from additional lenses. The CCD-detector is a low-etaloning \textsc{Pylon-100:BR-eXcelon}, $1340\times100$  pixels CCD, with a pixel size of $20\times20$\,$\mu$m$^2$. The quantum-efficiency lies above QE\textgreater90\% for the wavelength range $450$\,-\,$900$\,nm. 

\section{Experimental results}
\subsection{Optical phonon population and hole continuum dynamics in silicon}
Time-resolved Raman spectroscopy allows to address population dynamics of individual low-energy excitation modes after photo-excitation. Through detailed balance of the Stokes and anti-Stokes signals the population number of different modes can be determined, and calculated into effective mode temperatures.\cite{compaan1984} This direct way of measuring transient temperature evolution in the time-domain allows to test the validity and limitations of multiple effective temperature models,\cite{waldecker2016,maldonado2017} which on the shortest time-scales may not hold anymore due to the presence of truly non-thermal occupation statistics. Such effective temperature models are not only widely applied to describe ultrafast dynamics in simple materials, \cite{waldecker2016,maldonado2017} but also in quantum matter where different spin, orbital, electronic and lattice degrees of freedom are present with their respective excitations. \cite{averitt2001,ogasawara2005prl}

Raman tensor dynamics is a less treated aspect of time-resolved Raman scattering. It is however of crucial importance in the correct determination of effective mode temperatures. \cite{compaan1984} The anti-Stokes intensity scales as I$_{\rm AS}$\,$\propto$\,$\chi^2$\,$n$ , were $n$ is the mode occupation number, and $\chi^2$ the squared Raman tensor. The Stokes-intensity scales as I$_{\rm S}$\,$\propto$\,$\chi^2$\,$[n+1]$ and is thus less sensitive to transient population dynamics. This can be taken as a motivation to ''neglect'' dynamics of the Stokes-spectrum. However, from these relations it directly becomes clear that assigning transient anti-Stokes dynamics solely to transient population changes may not necessarily be true since the Raman tensor $\chi^2$ may also show photo-induced changes. This is expected to be of special importance under resonant probing conditions. We've recently demonstrated in silicon that for a correct transient mode temperature determination indeed a dynamical modification of the Raman tensor $\chi^2$ needs to be taken into account. \cite{zhu2017} The quench of $\chi^2$ in silicon was discussed to originate from the photo-induced hole density.  Below we reiterate our results on the LO-phonon population and hole continuum dynamics in silicon for similar measurement conditions. A dynamical modification of the Raman tensor is expected to play a role in many other materials such as $3d$/$4d$/$5d$-oxides, carbon allotropes \cite{ferrari2013raman} like graphene or Bucky-tubes, and transition metal dichalcogenides. \cite{lee2017}

An intrinsic (100) oriented silicon wafer (resistivity $>$\,$10000$\,$\Omega$m), at room temperature, is excited above the indirect band-gap with $650$\,nm pulses of $0.3$\,ps duration.  The Raman probe excitation is narrow-band ($\Delta\nu$\,$\approx$\,$10$\,cm$^{-1}$ FWHM) pulsed light of $512$\,nm central wavelength, and $1.5$\,ps temporal duration. The pump and probe powers were set to P$_{\rm pump}$\,=\,$12$\,mW and P$_{\rm probe}$\,=\,$1$\,mW for the pump and probe beam respectively. With a pump spot size of $100$\,$\mu$m diameter, the photo-excitation density is about $n_0$\,$\sim$\,$8$\,$\times$\,$10^{18}$\,cm$^{-3}$.  Both pump and probe beams are aligned with the [110] crystallographic directions. The backscattered light is collected with the single-stage spectrometer, where a notch-filter (\textsc{Kaiser Optical Systems Inc. SuperNotch-Plus} $514.5$\,nm) is used to block the elastically scattered light. 

\begin{figure}[h!]
\center
\includegraphics[scale=1]{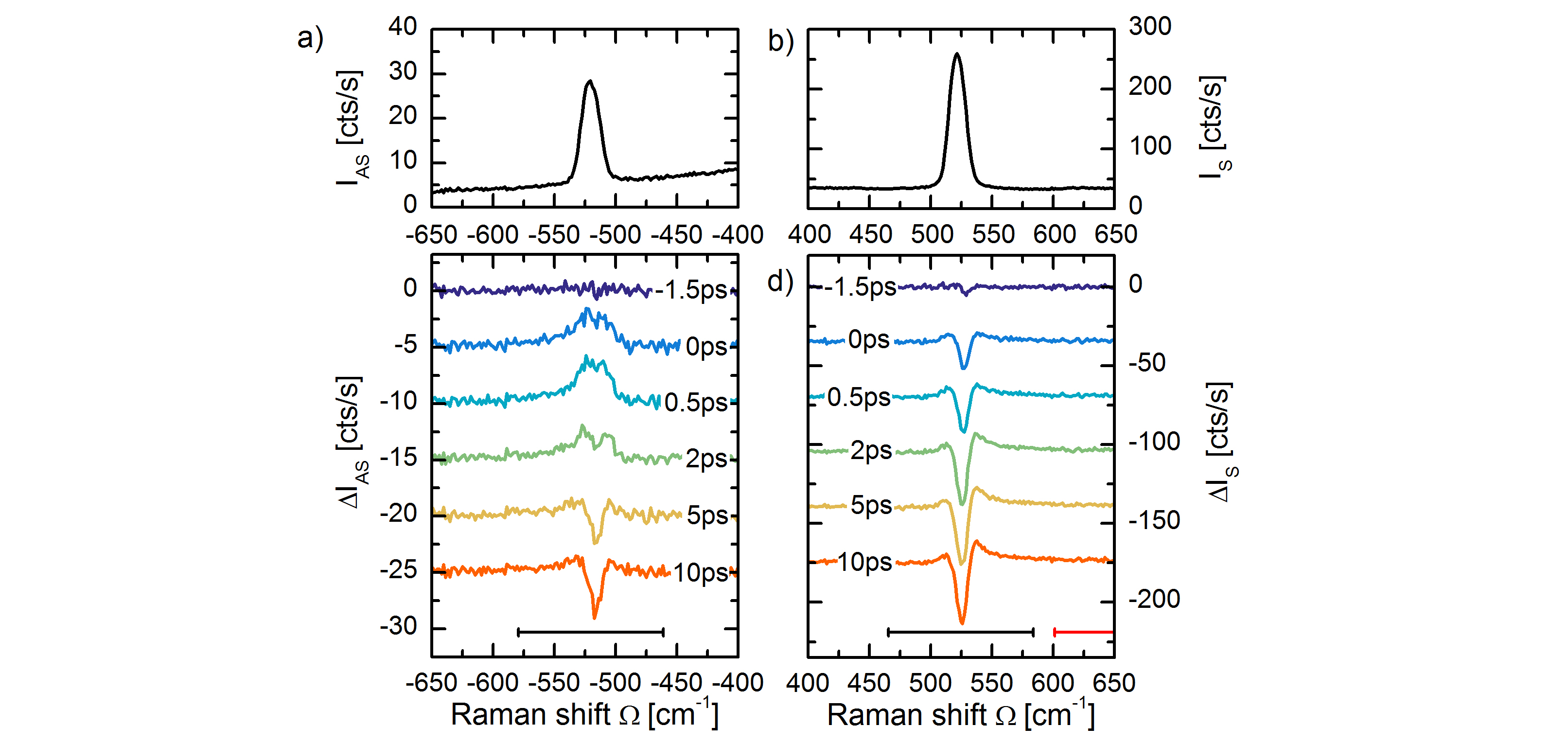}
\caption{a) Anti-Stokes I$_{\rm AS}$ and b) Stokes I$_{\rm S}$ spectra of silicon. The main feature is the LO-phonon at $\pm 520$\,cm$^{-1}$. The background is attributed to hole continuum scattering. c) Differential anti-Stokes $\Delta$I$_{\rm AS}$ and d) differential Stokes $\Delta$I$_{\rm S}$ spectra for various representative time-delays. The anti-Stokes intensity shows a positive transient phonon intensity at $0.5$\,ps and $2$\,ps, corresponding to the creation, and decay, of an LO-phonon population. The Stokes side shows a transient reduction in phonon scattering efficiency due to a Raman tensor quench.}
\label{fig:onephonon_1}
\end{figure}

\begin{figure}[h!]
\center
\includegraphics[scale=1]{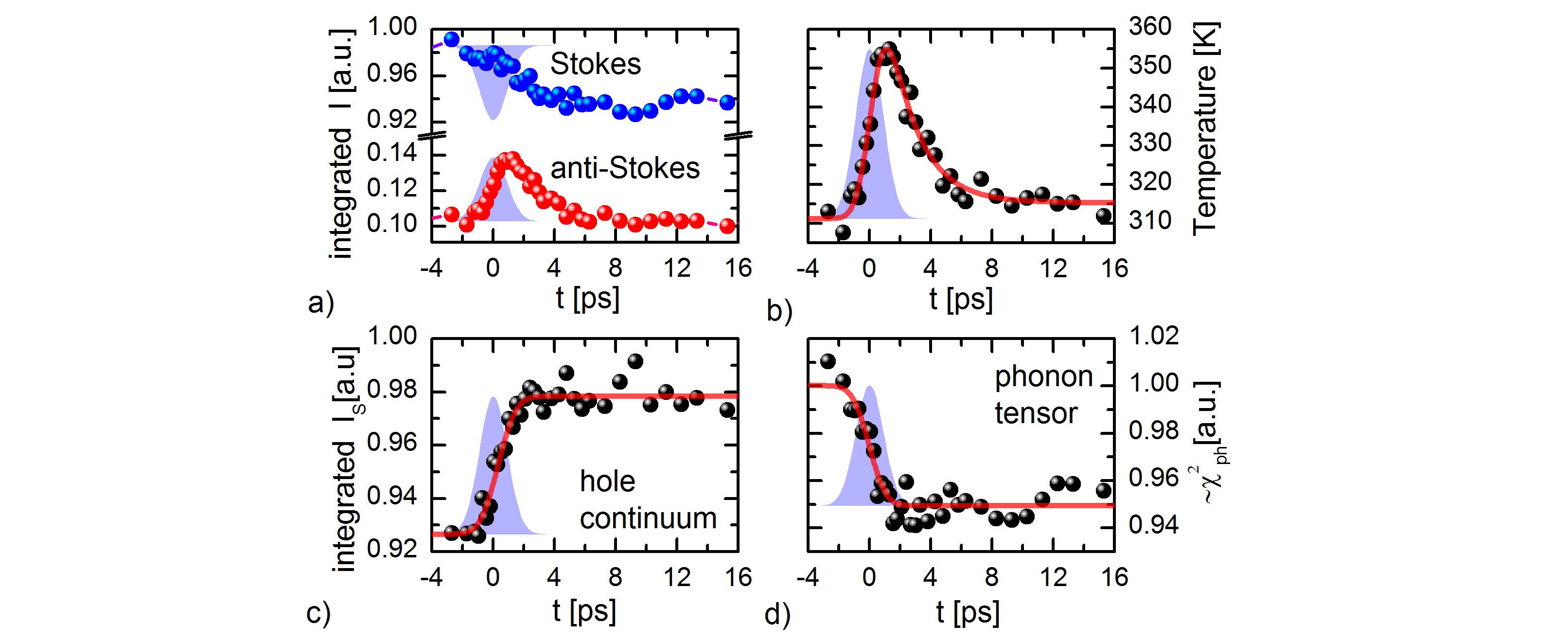}
\caption{a) Transients of Stokes and anti-Stokes phonon intensity, as integrated over the blue highlighted regions in the Stokes and anti-Stokes spectra. On the anti-Stokes side a positive transient intensity is observed, which corresponds to a transient optical phonon population. The dynamics on the Stokes side is dominated by the tensor quench. b) Transient phonon temperature calculated through detailed balance. The temperature rises from $310$\,K to about $355$\,K, and recovers on a time-scale of $\tau$\,$\approx$\,$2.0$\,$\pm$\,$0.6$\,ps. c) Transient hole continuum scattering on the Stokes side in the green highlighted region. d) A transient phonon tensor quench is observed. In all graphs the cross-correlation of the pump and probe pulse is plotted in faded blue.}
\label{fig:onephonon_2}
\end{figure}

Figure \ref{fig:onephonon_1}a and b show the anti-Stokes I$_{\rm AS}$ and Stokes I$_{\rm S}$ spectra of silicon (spectra at t\,=\,$-5$\,ps). The spectra consist of the zone-center longitudinal optical (LO) phonon at $\pm$\,$520$\,cm$^{-1}$, and an electronic continuum originating from hole scattering. \cite{hart1970,cerdeira1973} The silicon peak has a (Gaussian) half width at half maximum of $\sim$\,$8$\,cm$^{-1}$. This is larger than the intrinsic half-width of $\sim$\,$3$\,cm$^{-1}$ at room temperature, \cite{hart1970} since Gaussian ps-pulses were used to collect the Raman spectrum. In Fig.\,\ref{fig:onephonon_1}c and d the differential anti-Stokes scattering intensity $\Delta$I$_{\rm AS}$(t)\,=\,I$_{\rm AS}$(t)\,$-$\,I$_{\rm AS}{\rm (-5ps)}$, and Stokes scattering intensity $\Delta$I$_{\rm S}$(t)\,=\,I$_{\rm S}$(t)\,$-$\,I$_{\rm S}{\rm (-5ps)}$ is shown for various representative delay-times. At $0$\,ps to $0.5$\,ps a scattering increase is observed on the anti-Stokes side. This corresponds to the creation of a transient optical phonon population. However, on the Stokes side a negative differential feature is observed, which evidently cannot originate from a transient phonon population. In addition a small positive shoulder in the $\Delta$I$_{\rm S}$(t) spectra is observed. The integrated phonon scattering intensities are shown in Fig.\,\ref{fig:onephonon_2}a. The phonon response is integrated over the spectral regions $\pm$[$460$\,-\,$580$]\,cm$^{-1}$ (indicated with black bars) for the Stokes and anti-Stokes side respectively. A linear background was subtracted to take into account the increased hole continuum scattering over the integration region. The cross-correlation of the pump and probe pulse is plotted in faded blue for comparison.

The Stokes intensity I$_{\rm S}$, and anti-Stokes intensity I$_{\rm AS}$ are given as \cite{compaan1984}:

\begin{equation}
{\rm I}_{\rm S}(\omega_{\rm L}-\Omega)\propto[\omega_{\rm L}-\Omega]^3\,\times \,C(\omega_{\rm L}-\Omega,\omega_{\rm L})\,\times \,\chi^2(\omega_{\rm L}-\Omega,\omega_{\rm L})\,\times \,[n(\Omega)+1]
\label{eq:stokesintensity}
\end{equation}

\noindent and

\begin{equation}
{\rm I}_{\rm AS}(\omega_{\rm L}+\Omega)\propto[\omega_{\rm L}+\Omega]^3\,\times \,C(\omega_{\rm L}+\Omega,\omega_{\rm L})\,\times \,\chi^2(\omega_{\rm L}+\Omega,\omega_{\rm L})\,\times \,[n(\Omega)]
\end{equation}

Here $\chi^2$($\omega_{\rm L}$\,$\pm$\,$\Omega$,\,$\omega_{\rm L}$) gives the Raman tensor for the anti-Stokes and Stokes resonance, $\omega_{\rm L}$ is the probe excitation frequency, and $n(\Omega)$ is the occupation number for the mode with energy $\hbar\Omega$. The factors C($\omega_{\rm L}$\,$\pm$\,$\Omega$,\,$\omega_{\rm L}$) contain the incident intensity, and optical constants (absorption, transmission and refractive index) at the incident and scattered frequencies. \cite{compaan1984} We however neglect the frequency dependence of $C$ and $\chi^2$, which will lead to a small error of $<$\,$5$\,\% in the absolute phonon temperature determination. \cite{compaan1984}\cite{Note1} The anti-Stokes to Stokes scattering ratio can be used to calculate the phonon temperature, or equivalently the particle occupation number $n(\Omega)$, through the ratio:

\begin{equation}
\begin{split}
\frac{{\rm I}_{\rm AS}}{{\rm I}_{\rm S}}(\Omega) & =\frac{[\omega_{\rm L}+\Omega]^3}{[\omega_{\rm L}-\Omega]^3}\,\times\, \frac{C(\omega_{\rm L}+\Omega,\omega_{\rm L})}{C(\omega_{\rm L}-\Omega,\omega_{\rm L})}\, \times \,\frac{\chi^2(\omega_{\rm L}+\Omega,\omega_{\rm L})}{\chi^2(\omega_{\rm L}-\Omega,\omega_{\rm L})}\,\times \,\frac{n(\Omega)}{n(\Omega)+1}\\ & \approx\frac{[\omega_{\rm L}+\Omega]^3}{[\omega_{\rm L}-\Omega]^3}\times \exp(-\frac{{\rm \hbar} \Omega}{\rm k_BT})
\label{eq:intensityratio}
\end{split}
\end{equation}

In Fig.\,\ref{fig:onephonon_2}b the transient temperature, determined according to Eq.\,\ref{eq:intensityratio} is plotted. An average heating to $310$\,K is measured before time-zero. After photo-excitation the temperature rises by $\Delta$T\,$\approx$\,$45$\,K to $355$\,K within the time-resolution, followed by a decay with a time-scale of $\tau$\,$\approx$\,$2.0$\,$\pm$\,$0.6$\,ps. The rapid rise, and consecutive decay is consistent with the transient creation of an optical phonon population through electron-phonon coupling, which thereafter decays into acoustic phonons through anharmonic coupling. \cite{zhu2017,klemens1966} After $10$\,ps the temperature reaches quasi-equilibrium at an increased temperature of $\Delta$T\,$\approx$\,$5$\,K, which does not recover within the measured time window.

Figure \ref{fig:onephonon_2}c shows the transient increase of hole continuum scattering in the region +[$600$\,-\,$650$]\,cm$^{-1}$ (indicated with the red bar). The transient scattering signal is well-fitted with a step function convoluted with a cross-correlation function of the pump and probe pulse. The photo-induced electron-hole-density has other effects on the Raman spectrum. Since the phonon Raman transition probability is proportional to the density of electrons in the valence band, and holes in the conduction band, the photo-induced electron-hole density is expected to alter $\chi^2$. \cite{gillet2013,sangalli2016} From the Stokes intensity I$_{\rm S}$ and the calculated population $n$, the time-evolution of $\chi^2$ is calculated according to Eq.\,\ref{eq:stokesintensity}, and shown in Fig.\,\ref{fig:onephonon_2}d. A step-wise tensor quench $\Delta\chi^2$\,$\approx$\,$-5$\% is observed, underlining the electronic origin of the transient decrease of the Stokes intensity I$_{\rm S}$. The transient hole density in addition leads to an increase in the Fano-asymmetry, as evidenced by the ingrowing positive shoulder in the $\Delta$I$_{\rm S}$(t) spectra. The origin of the tensor quench and the transient Fano-asymmetry are discussed in more detail in Ref.\,\onlinecite{zhu2017}.

\subsection{Time- and momentum resolved scattering in silicon}
Electron-phonon, and phonon-phonon interaction strengths, or more generally the coupling between any type of quasiparticles, depends on the momentum of the interacting quasiparticles. \cite{birmanbook} The characteristic timescales associated with optical phonon creation, and relaxation through anharmonic coupling, are thus momentum-dependent. \cite{maldonado2017,tandon2015} Pump-probe techniques spanning from x-ray to the optical range allow for time- and momentum-resolved probing. Whereas the momentum-dependence is implicit for pump-probe techniques in the x-ray range, such as diffuse scattering, \cite{trigo2013fourier} diffraction, \cite{johnson2017} or resonant inelastic x-ray scattering, \cite{dean2016ultrafast} this may not directly be obvious for optical pump-probe techniques as these are $\mathbf{k}$\,$\approx$\,$0$ momentum transfer probes. Optical inelastic scattering techniques, notably femtosecond stimulated Raman spectroscopy, \cite{batignani2015probing} and spontaneous Raman spectroscopy however can be utilized to disentangle momentum-dependent dynamics through higher-order scattering processes. In a two-particle process, two excitations of opposite momenta $+\mathbf{k}$ and $-\mathbf{k}$ are created (or annihilated). \cite{cardonabook} Two-particle scattering processes appear in the inelastic light scattering spectrum as bands of the frequency dependence of the combined density of states of the excitation pair, weighted by a scattering efficiency distribution. \cite{johnson1964} The multi-particle scattering response thus forms a mapping of momentum-space to the energy domain, and thereby allows for the study of momentum-resolved particle dynamics. After careful analysis of the line-shape even momentum-relaxation dynamics should be in principle feasible to detect. \cite{lockwood1987} With time-resolved spontaneous Raman spectroscopy we succeeded to detect photo-induced phonon softening of different high-symmetry Brillouin Zone-points (BZ-points) in the two-phonon spectrum of silicon. We find evidence that the softening at the different points occurs with dissimilar time-scales, which is indicative of dissimilar electron-phonon scattering rates.

Probing of the two-phonon spectrum on a (100) oriented silicon wafer was performed at room temperature. A $512$\,nm Raman probe excitation was used, with the probe polarization along the [110] axis. The two-phonon spectrum is observed between $920$\,cm$^{-1}$ and $1040$\,cm$^{-1}$ as seen in Fig. 5a (dark blue line). The two-phonon spectrum consists of a summation of TO-phonon overtones from four high-symmetry Brillouin Zone (BZ) points. \cite{temple1973} Previous reports have assigned the characteristic spectral features based on the analysis of critical points in the phonon density-of-states. \cite{johnson1964,temple1973,gillet2017ab} The sharp increase at $920$\,cm$^{-1}$ is related to scattering from the X-point, the shoulder at $940$\,cm$^{-1}$ to the W-point, and the shoulder at $970$\,cm$^{-1}$ to scattering from the L-point. The tail extending from $980$\,cm$^{-1}$ to $1040$\,cm$^{-1}$ is the overtone from the zone center, i.\,e. the $\Gamma$-point. The different regions are marked with bars.

\begin{figure}[h]
\center
\includegraphics[scale=1]{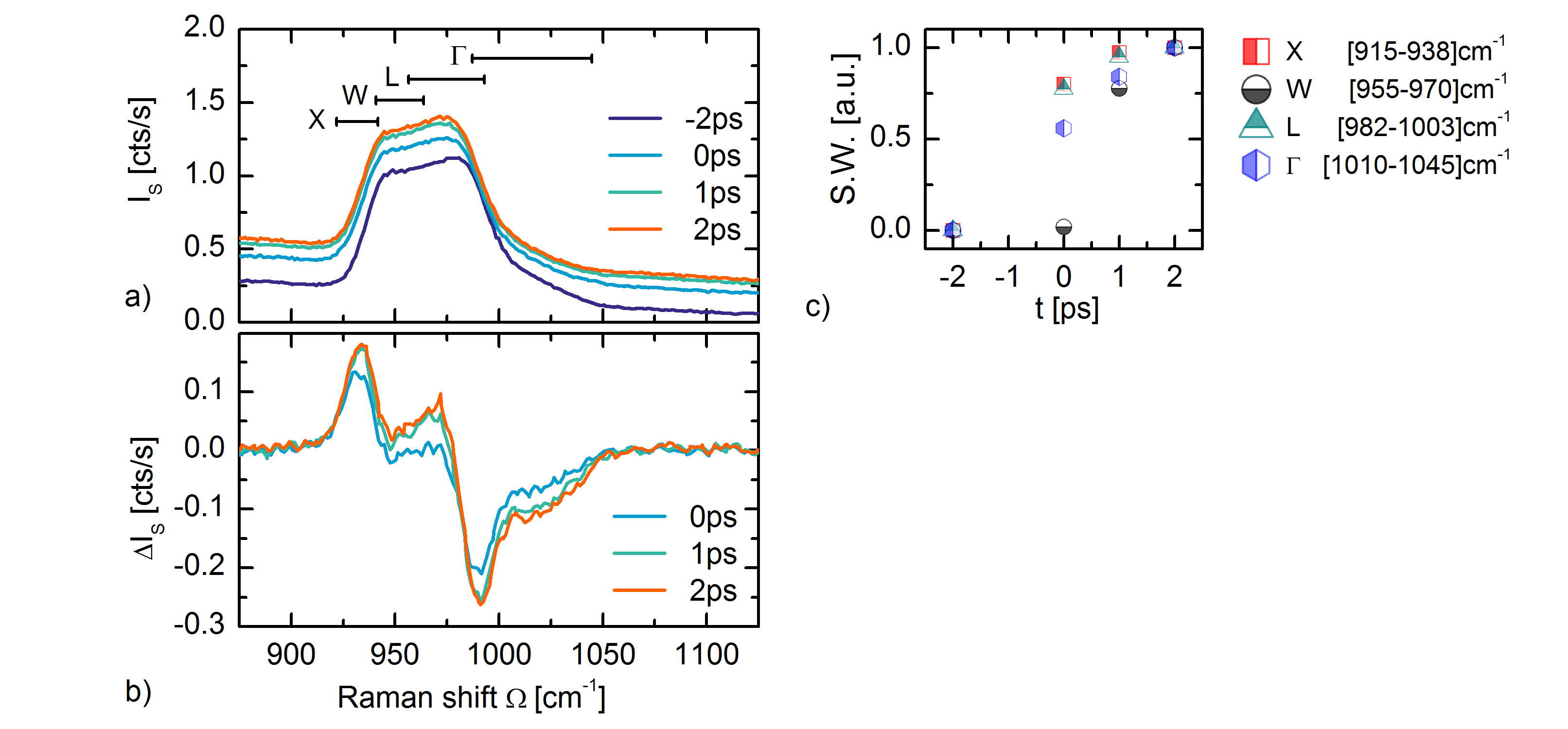}
\caption{a) Two-phonon spectra I$_{\rm S}$(t) of photo-excited Silicon for various delay-times. Scattering regions originating from the different Brillouin Zone regions X,W,L, and $\Gamma$ are indicated with bars. b) Corresponding differential spectra $\Delta$I$_{\rm S}$(t). The transient hole scattering response has been subtracted. The spectral shift at the W- and  $\Gamma$-point appears to have a slower ingrowth than the X- and L-point. c)  Spectral weight S.W. for representative regions in the two-phonon signal normalized on the S.W. at $+2$\,ps.}
\label{fig:twophonon}
\end{figure}

The sample is excited above the indirect band-gap with $650$\,nm pulses of $0.3$\,ps duration, and pump polarization along the [110] axis. The photo-induced electron-hole density is $n_0$\,$\sim 9$\,$\times$\,$10^{19}$\,cm$^{-3}$. Figure \ref{fig:twophonon}a shows the transient Stokes spectra I$_{\rm S}$(t) of the two-phonon peak for various pump-probe delays. A photo-induced redshift of the two-phonon spectrum is observed, in addition to an increased hole continuum scattering background, as discussed above. Figure \ref{fig:twophonon}b shows the differential spectra $\Delta$I$_{\rm S}$(t)\,=\,I$_{\rm S}$(t)\,$-$\,I$_{\rm S}$($\rm -2ps$) for the respective time-delays, where the hole continuum scattering increase is subtracted (the tail of a Voight profile is fitted to the increase in electronic scattering in the $\Delta$I$_{\rm S}$(t) spectra). The $\Delta$I$_{\rm S}$(t)\,/\,I$_{\rm -2ps}$ spectra are integrated over four different wavelength regions and plotted normalized to the I$_{\rm S}$\,/\,I$_{\rm -2ps}$ value at $+2$\,ps. This gives the spectral weight S.W. as a function of time delay as shown in Fig.\,\ref{fig:twophonon}c.

The two-phonon Stokes signal has a I$_{\rm S}(\omega)$\,$\propto$\,$\chi^2(\omega) [n^2(\Omega)+1]$ dependence (cf.\,Eq.\,\ref{eq:stokesintensity}). Both the Raman tensor $\chi^2(\omega)$ and the population factor $[n^2(\Omega)+1]$ transiently change upon photo-excitation as discussed above, where the population term can only increase. The tensor dynamics $\chi^2(\omega)$ thus appears to dominate the transient two-phonon response.  This is also evidenced by the large redshift $\Delta\Omega$\,$\approx$\,$-2$\,cm$^{-1}$. The redshift is understood as a photo-induced phonon softening at the different high symmetry points in the Brillouin Zone. Extracting time-constants for the different BZ-points is not possible due the limited amount of time-delay points. However, from the $\Delta$I$_{\rm S}$(t) spectra in Fig.\,\ref{fig:twophonon}b, and the spectral weight S.W. in Fig.\,\ref{fig:twophonon}c, it appears that the phonon softening at the X- and L-point has a faster ingrowth than at the W- and $\Gamma$\,-point. This hints to dissimilar electron-phonon coupling constants for the different points. The quantitative determination of time-constants for the four BZ-points asks for a higher time-resolution. In spontaneous Raman spectroscopy this would however result in a loss of frequency resolving power, and thereby momentum-resolving power. This would again lead to a complication in disentangling time-and momentum resolved scattering, however now because of frequency resolution limitations. A qualitative disentanglement of momentum-dependent scattering in silicon thus forms an interesting study case for the time-resolved femtosecond stimulated Raman scattering variant. 

\texorpdfstring{\subsection{Photo-induced melting of helimagnetic order in the chiral magnet Cu$_2$OSeO$_3$}}

The understanding of out-of-equilibrium phenomena in magnetic materials is a highly active part of condensed matter research. \cite{kirilyuk2010} Recent scientific advances with a high potential towards applications include all-optical magnetization switching, \cite{savoini2012} and picosecond optical writing and read-out of magnetic bits. \cite{stupakiewicz2017ultrafast} With increasingly complex magnetic phases and dynamical phenomena being studied, there is a high demand for novel all-optical probes.  Magnetization dynamics in finite magnetization phases, such as ferromagnetic and ferrimagnetic order, or a more exotic phase such as the skyrmion lattice, \cite{ogawa2015ultrafast} can be probed by optical techniques which are sensitive to the order parameter \textbf{M} which describes the (macroscopic) magnetization. Linear magneto-optical effects, for instance the Faraday or magneto-optical Kerr effect, measure a polarization rotation $\theta$\,$\propto$\,$\mathbf{M}$ proportional to the magnetization. \cite{pershan1967magneto} The dynamical variant allows to detect photo-induced incoherent and coherent dynamics of the order parameter \textbf{M}. \citep{ogasawara2005prl,ogawa2015ultrafast} All-optical detection of antiferromagnetic order dynamics already becomes more challenging. The antiferromagnetic order parameter \textbf{L} is proportional to the difference in magnetic sublattice magnetizations; $\mathbf{L}$\,$\propto$\,$\mathbf{M}_{\uparrow}$\,-\,$\mathbf{M}_{\downarrow}$. Optical probing of \textbf{L} is realized by second-order magneto-optical effects, such as magnetic linear birefringence and second harmonic generation. Ultrafast variants have been used to study photo-induced dynamics in antiferromagnets. \cite{bossini2014controlling,fiebig2008ultrafast} For more complex net-zero magnetization order, such as cycloidal, helical, or spin wave order it is more challenging to probe the magnetic order parameter.  One suggestion is tracking the thermalization of a magnon mode with THz-time-domain spectroscopy after photo-excitation, as was realized in the spin-cycloid material TbMnO$_3$. \cite{bowlan2016} The Raman-activity of a magnon mode would however permit all-optical detection, as opposed to detection through its dipole activity. Here we show that time-resolved Raman spectroscopy allows to all-optically probe melting of helimagnetic order after photo-excitation in the chiral magnet Cu$_2$OSeO$_3$. \cite{seki2012} This is realized by tracking the softening and broadening of a Raman-active magnon mode in the time-domain.

\begin{figure}[h]
\center
\includegraphics[scale=1]{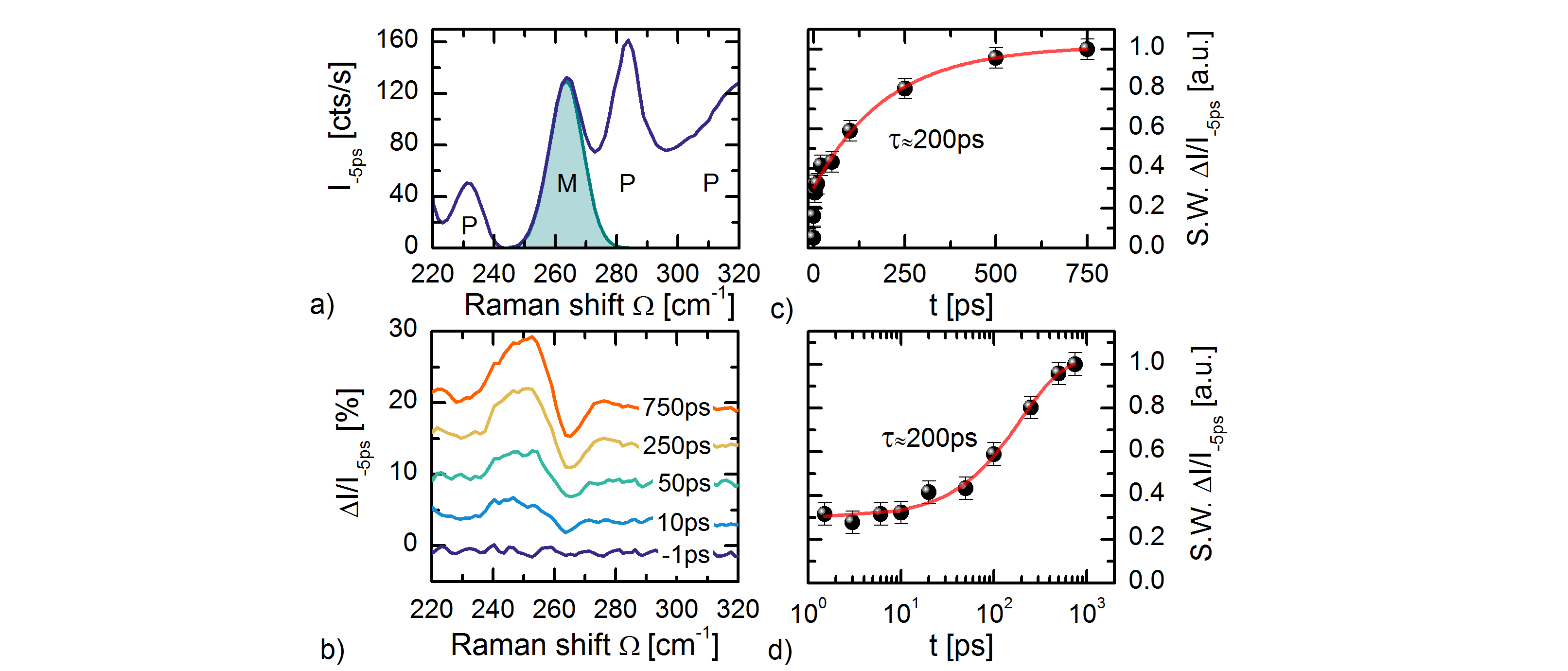}
\caption{a) Part of the Raman spectrum of Cu$_2$OSeO$_3$ recorded with $512$\,nm pulsed probe light. The peak labeled ''M'' at $\Omega_{\rm M}$\,$\approx$\,$264$\,cm$^{-1}$ is a  $\Delta S^z$\,=\,$+1$ spin cluster magnon. The peaks indicated by ''P'' are phonons, or regions of not fully resolved phonons. b) Scaled differential Stokes spectra $\Delta$I$_{\rm S}$(t)\,/\,I$_{\rm -5ps}$ for representative delay-times. The derivative-like lineshape results from a dynamical softening and broadening of the magnon peak. c) linear and d) logarithmic scales of the spectral weight (S.W.) evolution of the ingrowing component in the $\Delta$I$_{\rm S}$(t)\,/\,I$_{\rm -5ps}$ spectra. The spectral weight transfer has a time-constant of $\tau$\,$\approx$\,$200$\,ps.}
\label{fig:magnon}
\end{figure}

Cu$_2$OSeO$_3$ has a long-range magnetic ordering temperature of T$_{\rm N}$\,$\approx$\,$58$\,K. \cite{seki2012} The magnetic ground state consists of helimagnetically aligned effective $S$\,=\,$1$ spin clusters, which by itself consist of three-up-one-down $S$\,=\,$1/2$ spins on the Cu$^{2+}$($3d^9$) sites. \cite{janson2014quantum,ozerov2014} The spin cluster ordering results in high energy $\sim$\,meV magnons, from which the magnons at the $\Gamma$-point are observable by spontaneous Raman spectroscopy. \cite{gnezdilov2010} A [111] oriented Cu$_2$OSeO$_3$ sample, at bias temperature $5$\,K in the helimagnetic phase was studied. The SHBC output at $512$\,nm, with $\Delta\nu$\,$\approx$\,$10$\,cm$^{-1}$ FWHM was used as Raman probe ($0.5$\,mW probe power at sample position). In Fig.\ref{fig:magnon}a part of the Stokes spectrum is shown. The M-mode at $\Omega$\,$\approx$\,$264$\,cm$^{-1}$ corresponds to a $\Delta S^z$\,=\,$+1$ magnon mode, which can be understood as a spin-flip excitation within the three-up-one-down spin cluster. \cite{romhanyi2014} The modes labeled ''P'' correspond to phonons, or regions of not fully resolved overlapping phonons.  

Local crystal field excitations \cite{versteeg2016} were excited with $0.3$\,ps, $\lambda$\,=\,$570$\,nm pump pulses (with $2$\,mW pump power, and fluence F\,$\approx$\,$2$mJ/cm$^2$), after which the thermalization dynamics is measured in the transient Raman spectra. In Fig.\,\ref{fig:magnon}b the scaled differential spectra $\Delta$I$_{\rm S}$(t)\,/\,I$_{\rm -5ps}$ are shown for representative time-delays, where $\Delta$I$_{\rm S}$(t)\,=\,I$_{\rm S}$(t)\,$-$\,I$_{\rm S}$($\rm -5ps$). A derivative-like lineshape is observed in the $\Delta$I$_{\rm S}$(t)\,/\,I$_{\rm -5ps}$ spectra, resulting from a spectral weight shift to lower Raman shift $\Omega$ for the magnon excitation. For the low temperatures, and Stokes range of interest ($\Omega$\,=$220$-$320$\,cm$^{-1}$) the occupation number $n$($\Omega$,T)\,$\ll$\,$1$. The change in occupation thus also is $\Delta n$($\Omega$,T)\,$\ll$\,$1$. We thereby assign the spectral weight transfer to a change in the Raman tensor $\chi ^2(\omega)$. This physically corresponds to a dynamic softening and broadening of the magnon excitation. From fitting of the $\Delta S^z$\,=\,$+1$ peak position in the I$_{\rm S}$(t) spectra it is found that the dynamic shift $\Delta \Omega$($750$\,ps) at late delay times lies below $1$\,cm$^{-1}$. By comparison to the temperature dependent position we conclude that the transient magnetic temperature does not rise above $25$\,K. Under our pump excitation conditions the long-range magnetic order thus gets strongly perturbed, but not fully destroyed. The phonon peak shifts are too small to be resolvable in the $\Delta$I$_{\rm S}$(t)\,/\,I$_{\rm -5ps}$ spectra. The ingrowing spectral weight transfer (S.W.) of the $\Delta$I$_{\rm S}$(t)\,/\,I$_{\rm -5ps}$ signal, corresponding to the temporal evolution of the magnetic order parameter, is shown in Fig.\,\ref{fig:magnon}c and d in linear and logarithmic timescales. A typical time-scale of $\tau$\,$\approx$\,$200$\,ps for spin-lattice thermalization in insulators is observed. \cite{kirilyuk2010} This agrees well with the findings of \textit{Langner et al.} where a spin-lattice thermalization time of $\tau$\,$\approx$\,$300$\,ps was observed by means of time-resolved resonant x-ray diffraction. \cite{langner2017} The slight discrepancy might originate from the different pump excitation conditions, and the different bias temperatures.

\section{Conclusions and outlook}
In this paper have presented a flexible and efficient ultrafast time-resolved spontaneous Raman spectroscopy setup, and illustrated its strength and capabilities with different conceptual time-resolved Raman studies of collective excitation and quasi-particle dynamics. One of the strongest feats of time-resolved spontaneous Raman is the determination of transient population numbers, and effective modes temperatures through detailed balance of the anti-Stokes to Stokes intensity ratio. A less studied aspect of the dynamics is the resonance enhancement in the Raman process. This however plays a role, as we demonstrated in the prototype test material silicon. We've shown that the photo-induced hole density leads to a quench of the Raman tensor, and in addition to an increase in electronic inelastic scattering. Dynamic changes in the electronic population and structure are of importance to photo-induced phenomena in many classes of quantum matter, such as $3d$/$4d$/$5d$-materials, carbon allotropes, and transition metal dichalgocenides. Time-resolved resonant Raman can thereby provide a convenient tool to study both electronic population dynamics, as well as photo-induced changes in the electronic band structure such as band gap renormalization. 

Although Raman spectroscopy is a zero-momentum transfer probe, momentum-dependent scattering can nevertheless be resolved in the higher-order Raman response. The overtone spectrum can thus serve as a momentum-and time-resolved scattering probe. As a proof of concept, we've studied photo-induced changes in the two-phonon response of silicon, and found evidence for dissimilar phonon softening rates at different high-symmetry points of the Brillouin Zone. Applying this concept to slower phenomena, such as spin-lattice relaxation in $3d$-antiferromagnetic materials, seems fortuitous to pursue, and possibly allows to detect momentum-dependent transient population dynamics and momentum relaxation.

Raman-active collective excitations such as magnons, charge-density wave modes, and the Cooper-pair breaking peak, can serve as a time-resolved probe for the order parameter of quantum phases. We illustrated this in the helimagnet Cu$_2$OSeO$_3$, where the photo-induced melting of helimagnetic order can be tracked by the observation of a dynamic softening and broadening of a magnon peak. This principle can also be applied to probe dynamics of other net-zero magnetization order such as cycloidal, spin-density wave, and antiferromagnetic order. On a broader scope this can be applied to photo-induced phase transitions, where the associated change of symmetry can be derived from the transient Raman spectrum. The Raman spectrum of collective low energy excitations can additionally serve as a probe to study energy and angular momentum transfer between the lattice, electronic, orbital and magnetic degrees of freedom in quantum matter. 

\section{Acknowledgements}
This project was partially financed by the Deutsche Forschungsgemeinschaft (DFG) through SFB Grossger\"{a}teantrag INST217/782-1, and SFB-1238 (projects A02 and B05). R.B.V. acknowledges funding through the Bonn-Cologne Graduate School of Physics and Astronomy (BCGS).

\end{document}